\documentclass[prb,aps,twocolumn,superscriptaddress]{revtex4-1}

\usepackage{graphicx}
\usepackage[english]{babel}
\usepackage{amsmath}

\usepackage{bm}              
\usepackage{w-greek}         
\usepackage{graphicx}
\usepackage{units}

\usepackage{ulem}
\usepackage{color}

\begin{document}

\title{Simulations on the influence of spatially-varying spin transport parameters on the measured spin lifetime in graphene non-local spin valves}

\author{Marc Dr\"ogeler}
\author{Frank Volmer}
\affiliation{2nd Institute of Physics and JARA-FIT, RWTH Aachen University, D-52074 Aachen, Germany}
\author{Christoph Stampfer}
\affiliation{2nd Institute of Physics and JARA-FIT, RWTH Aachen University, D-52074 Aachen, Germany}
\affiliation{Peter Gr\"unberg Institute (PGI-9), Forschungszentrum J\"ulich, 52425 J\"ulich, Germany}
\author{Bernd Beschoten}
\thanks{E-mail address: bernd.beschoten@physik.rwth-aachen.de}
\affiliation{2nd Institute of Physics and JARA-FIT, RWTH Aachen University, D-52074 Aachen, Germany}

\date{\today}

\begin{abstract}

Spin transport properties of graphene non-local spin valve devices are typically determined from Hanle spin precession measurements by using a simplified solution of the one-dimensional Bloch-Torrey equation which assumes infinitely long transport channels and uniform spin transport parameter. We investigate the effects of a finite graphene size and explore the influence of spatially-varying transport parameters on the measured Hanle curves by finite element simulations. We assume enhanced spin dephasing in the contact-covered graphene areas with additional Fermi level pinning and explore the influence of non-magnetic reference electrodes which are not properly decoupled from graphene. In experiments, it is typically observed that the spin lifetime increases with increasing charge carrier density. None of our simulations can reproduce this trend indicating that this dependency originates from spin transport through graphene areas which are not covered by contacts. We find that the extracted spin lifetime might be overestimated in flakes which are shorter than the spin diffusion length. Moreover, contact-induced spin dephasing leads to an overall reduction of the extracted spin lifetime. Additionally, we show that non-magnetic reference electrodes may also influence the measured spin lifetime even though they are not part of the transport area under investigation.

\end{abstract}

\maketitle


\maketitle 

\section{Introduction}

Graphene is one of the most promising materials in the field of spintronics as it combines long spin lifetimes and long spin diffusion length which results from high charge carrier mobilities at room temperature \cite{PhysRevB.74.155426,Pesin2012,PhysRevB.80.041405,GrapheneSpintronics,2053-1583-2-3-030202}. In most experiments the spin transport parameters such as the spin lifetime and the spin diffusion length are extracted from Hanle spin precession measurements using non-local graphene-based spin valve devices \cite{Tombros2007}. The resulting Hanle depolarization curves are often fitted by a simplified solution of the one-dimensional Bloch-Torrey equation \cite{PhysRev.104.563,PhysRevB.37.5312,ISI000249789600001}. This solution assumes that the graphene flake is infinitely long, i.e. much longer than the spin diffusion length, and that the spin transport parameter are uniform across the whole flake.

In experiments, however, the flake size is typically restricted to a few tens of micrometers. Due to recent improvements in device quality, the spin diffusion length now approaches the overall sample size \cite{Droegeler2014,doi:10.1021/acs.nanolett.6b00497,PhysRevLett.113.086602,PhysRevB.92.201410,doi:10.1021/acs.nanolett.6b01004,doi:10.1063/1.4962635,Kamalakar2015}. Additionally, confocal Raman measurements revealed that there are strong doping variations along the graphene transport channel, i.e. graphene areas in direct contact to spin injection and spin detection electrodes are typically highly doped with doping densities exceeding $10^{12}\text{cm}^{-2}$, which is more than one order of magnitude larger than for contact-free graphene areas \cite{Droegeler2014}. Moreover, there is strong evidence that the spin lifetime is reduced in the area of the electrodes whenever there is interaction between the metal and graphene \cite{Kamalakar2015,PhysRevB.88.161405,PhysRevB.90.165403}. Hence, the assumption of uniform spin transport parameters may not be valid and, therefore, it is interesting to investigate the effect of spatially-varying transport parameter. In addition, it is typically observed that the spin lifetime increases with increasing gate voltage, i.e. increasing charge carrier density \cite{Droegeler2014,doi:10.1021/acs.nanolett.6b00497,PhysRevLett.113.086602,PhysRevB.88.161405,PhysRevB.90.165403,PhysRevB.80.241403,PhysRevLett.107.047207,doi:10.1021/nl301567n,PhysRevLett.107.047206,Avsar2011,PSSB201552418}. Although there are several theories which can explain such a behavior \cite{Tuan2014,PhysRevB.94.041405,PhysRevLett.112.116602}, none of them can unambiguously be proven up to now. Hence, it is interesting to explore if spatially-varying transport parameters can also lead to the observed gate dependence.


In this paper we use a finite element simulation to investigate the influences of both a finite flake size and spatially-varying spin transport parameters on the measured Hanle curves. In the first part (Section~3) we investigate the effect of a finite flake size where uniform transport parameter are considered. In the second part (Section~4) we look at spatially-varying spin transport parameter caused by the electrodes themselves, while in a last part (Section~5) we focus on the impact of metallic reference electrodes which are not part of the actual spin transport channel.

\section{Simulation details}

\begin{figure*}[tbh]
	\centering
		\includegraphics{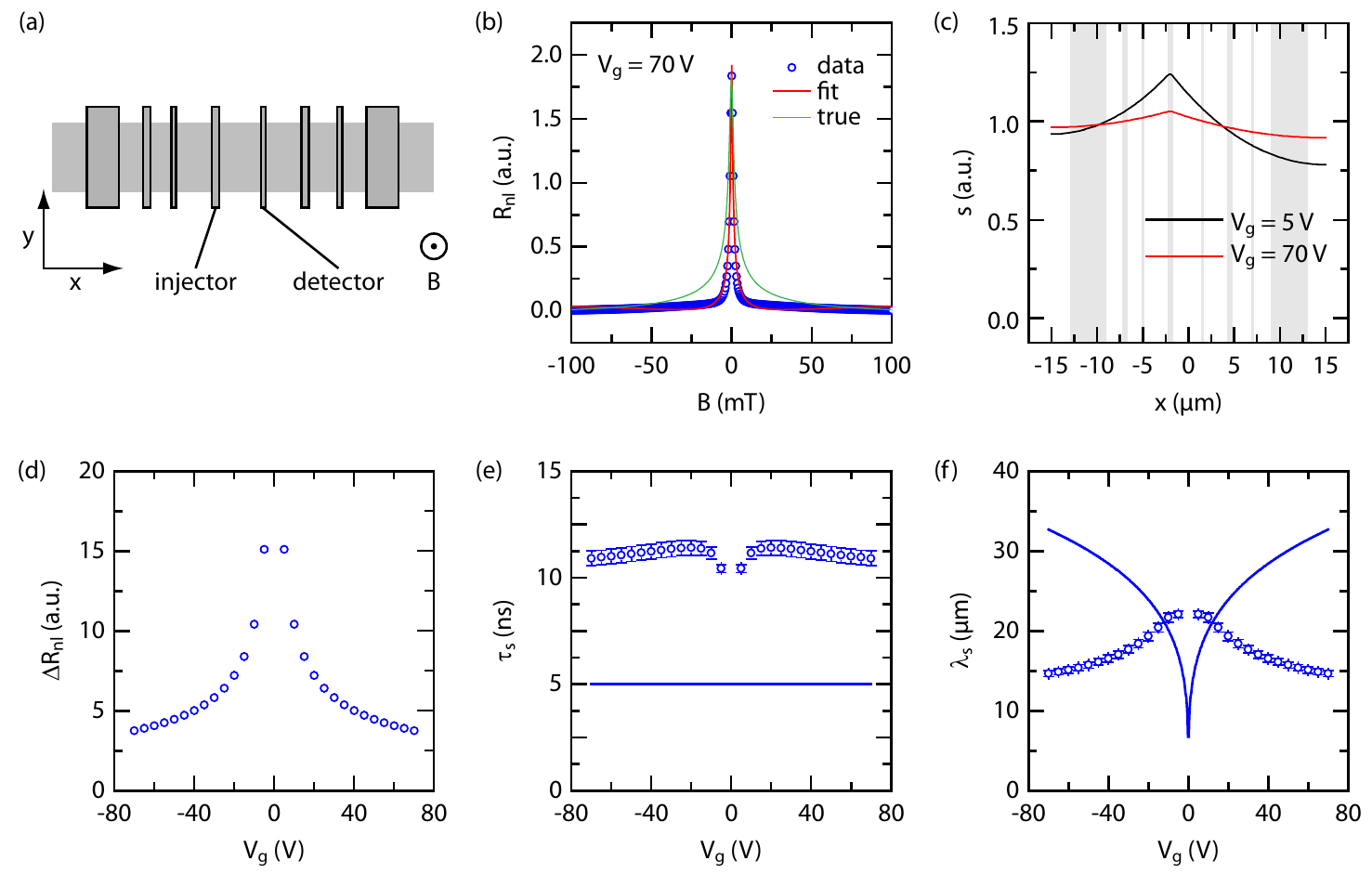}
	\caption{Simulation of a graphene flake with uniform spin transport parameters with $\tau_\text{s}=\unit[5]{ns}$ and $\mu=\unit[20,000]{cm^2/(Vs)}$. (a) Geometry of the spin transport device with assigned spin injection and spin detection electrodes. (b) Simulated Hanle curve for $V_\text{g}=\unit[70]{V}$. Next to the simulated data points also the fit (red curve) and the expected Hanle curve (green curve) for the genuine spin transport parameters are included. (c) Spatial variation of the spin accumulation for two back gate voltages $V_\text{g} = \unit[5]{V}$ and $V_\text{g} = \unit[70]{V}$. The positions of the contacts are indicated by the grey areas. (d-f) Fitted spin transport parameters as a function of gate voltage: (d) The magnitude of the spin signal $\Delta R_\text{nl}$, (e) the spin lifetime $\tau_\text{s}$, and (f) the spin diffusion length $\lambda_\text{s}$. All data were extracted from the simulated Hanle curves. The solid lines in (e) and (f) show the respective values which were used as the parameters for the simulations.}
	\label{fig:Sim_LongLifetimes}
\end{figure*}

For the simulations we use a finite element solver (COMSOL Multiphysics$^\text{\textregistered}$) which simulates the evolution of an injected spin accumulation in two dimensions. The size of the graphene flake is assumed to be $\unit[30]{\mu m} \times \unit[5]{\mu m}$ which corresponds to a typical flake size used in experiments. A sketch of the device structure is shown in Figure~\ref{fig:Sim_LongLifetimes}a. The electrode configuration is comparable to the layout used in some of our previous spin transport experiments \cite{Droegeler2014,doi:10.1021/acs.nanolett.6b00497,PSSB201552418}. The electrode width for the outer contacts reference electrodes is $W=\unit[1]{\mu m}$ and is alternating between $W=\unit[600]{nm}$ and $W=\unit[300]{nm}$ for the inner electrodes. The distance between electrodes varies between $L=\unit[2]{\mu m}$ and $L=\unit[3.5]{\mu m}$. For the sake of simplicity we assume a constant and uniform spin injection rate throughout the whole area of the spin injection contact.
Interface effects such as spin dependent back scattering \cite{PhysRevB.86.235408,PhysRevB.91.241407} into the electrodes are neglected. For the spatial evolution of the local injected spin accumulation $\pmb{s}(x,y)$ we use the Bloch-Torrey equation~\cite{PhysRev.104.563,PhysRevB.37.5312,ISI000249789600001}
\begin{equation}
\frac{\partial \pmb{s}(x,y)}{\partial t}\;=\;\pmb{s}(x,y)\times \pmb{\omega}+D_{\text{s}}\nabla^2\pmb{s}(x,y)-\frac{\pmb{s}(x,y)}{\tau_{\text{s}}}+\pmb{f}(x),
\label{eq:}
\end{equation}
where $\pmb{\omega} = \frac{g \mu_\text{B} \pmb{B}}{\hbar}$ is the Larmor frequency with the Bohr magneton $\mu_\text{B}$, the Land\'{e} factor $g=2$ for graphene and the magnetic field $\pmb{B}$. Furthermore, $D_\text{s}$ is the spin diffusion coefficient and $\tau_\text{s}$ the spin lifetime. From these values the spin diffusion length $\lambda_\text{s} = \sqrt{D_\text{s}\tau_\text{s}}$ can easily be calculated. Additionally, $\pmb{f}(x)$ is the spin injection rate with $\pmb{f}(x)\neq\pmb{0}$ in the area of the injection electrode and $\pmb{f}(x)=\pmb{0}$ elsewhere. We note that we do not consider spin drift effects \cite{doi:10.1021/acs.nanolett.6b01004,PhysRevB.79.081402} and solely focus on spin diffusion towards the spin detector. The spin diffusion coefficient $D_\text{s}$ is assumed to be equal to the charge diffusion coefficient $D_\text{c}$ which can be calculated from the charge carrier mobility $\mu$ and the induced charge carrier density $n$ via the conductance $\sigma = ne\mu$ with $\sigma = e^2 \nu(E_\text{F}) D_\text{c}$ \cite{RevModPhys.83.407}. Here, $e$ is the electron charge and $\nu(E_\text{F})$ the density of states in graphene at the Fermi energy. Moreover, the charge carrier density is expressed via the applied gate voltage $V_\text{g}$ using a standard capacitance model as $n = \alpha V_\text{g}$ where $\alpha=\unit[7.18 \times 10^{14}]{/(m^2 V)}$ is the lever arm for $\unit[300]{nm}$ SiO$_2$ \cite{Novoselov2004}. We assume the charge neutrality point to be at $V_\text{g}=\unit[0]{V}$ if not stated otherwise. Hence, the spin lifetime, the charge carrier mobility and the applied gate voltage are input parameters for the simulation. All parameters can additionally be varied and/or fixed for different regions of the device. The simulation calculates the steady-state spin distribution within the graphene flake for both different gate voltages (see Figure~\ref{fig:Sim_LongLifetimes}c) and different out-of-plane magnetic fields $\pmb{B}=B\pmb{e}_\text{z}$. The latter is used to evaluate the spin properties at the detection electrode only where we integrate the resulting spin accumulation in y-direction over the whole contact area. This approach assumes an ideal reference electrode in the experiment which is either non-magnetic or is placed at a distance relative to the injector where the spin accumulation is completely decayed. When plotting the integrated spin accumulation as a function of the applied magnetic field we reproduce the symmetric Hanle curve from our simulation (see Figure~\ref{fig:Sim_LongLifetimes}b) which we fit by the standard Hanle function (see Ref.~\cite{PhysRevB.90.165403}).

\section{Influence of flake sizes}

First we investigate the influence of the graphene flake size on the spin transport parameters. For this we assume a charge carrier mobility of $\mu=\unit[20,000]{cm^2/(Vs)}$ and a spin lifetime of $\tau_\text{s}=\unit[5]{ns}$ which are typical values recently measured on advanced spin valve devices at room temperature \cite{Droegeler2014,doi:10.1021/acs.nanolett.6b00497,PhysRevLett.113.086602,PhysRevB.92.201410}. All parameters are assumed to be uniform across the whole graphene flake. The extracted gate dependent spin transport parameters are shown in Figures~\ref{fig:Sim_LongLifetimes}d-f. The amplitude of the spin signal $\Delta R_\text{nl}$ in Figure~\ref{fig:Sim_LongLifetimes}d is given by the difference in non-local resistance between parallel and antiparallel alignments of the respective magnetization directions of the spin injection and detection electrodes at $B=\unit[0]{T}$. It shows a decrease with increasing gate voltage as expected from theory \cite{PhysRevLett.105.167202}. Due to an enhanced spin diffusion coefficient at large gate voltages the spins can diffuse further and the resulting decay of the spin accumulation diminishes and, hence, $\Delta R_\text{nl}$ gets smaller. The spin lifetime in Figure~\ref{fig:Sim_LongLifetimes}e shows an M-shape with a weak decrease of $\tau_\text{s}$ towards $V_\text{g}=\unit[0]{V}$. Most importantly, the extracted spin lifetime exceeds $\unit[10]{ns}$ and is thus overestimated by more than a factor of two compared to the true spin lifetime of $\tau_\text{s}=\unit[5]{ns}$ (see blue solid line in Figure~\ref{fig:Sim_LongLifetimes}e). This effect is in accordance to a previous simulation by Wojtaszek et al. \cite{PhysRevB.89.245427}. The spin diffusion length in Figure~\ref{fig:Sim_LongLifetimes}f exhibit a qualitatively different gate dependence (open circles) than the expected V-shape which results from the constant spin lifetime combined with a gate dependent increase of the spin diffusion coefficient (solid line). The decrease of the extracted $\lambda_\text{s}$ values towards larger gate voltages indicates that the spin diffusion process gets suppressed. This behavior can be understood from the gate dependent decay of the magnitude of the spin accumulation $s$ along the graphene channel which is plotted in Figure~\ref{fig:Sim_LongLifetimes}c for $V_\text{g}=\unit[5]{V}$ and $V_\text{g}=\unit[70]{V}$ at $B=\unit[0]{T}$. It can be seen that the spins can reach the end of the graphene flake at $x=\unit[\pm 15]{\mu m}$ where they can be reflected and once again reach the detection electrode. Due to their longer path length these spins acquire a larger phase than spins which directly reach the detector electrode without being reflected. Hence, the overall spin accumulation underneath the detector dephases faster with increasing magnetic field when reflection at the end of the graphene flake becomes relevant. This leads to narrower Hanle curves which mimics longer spin lifetimes. This effect can be seen in Figure~\ref{fig:Sim_LongLifetimes}b where the simulated Hanle curve for $V_\text{g}=\unit[70]{V}$ is shown. Additionally, the respective fit (red curve) and the expected Hanle curve (green curve) are included. This explains why the analysis of the Hanle curves overestimates the extracted spin lifetime.

When comparing the simulations to the experimental data, especially to the device which exhibit a spin lifetime of $\tau_\text{s}=\unit[12.4]{ns}$ (see Ref.~\cite{doi:10.1021/acs.nanolett.6b00497}), it might be assumed that the long spin lifetimes may result from the reflection at the ends of a short graphene flake. In this case the actual spin lifetime would be shorter. However, the device in question clearly shows the expected increase in spin diffusion lengths with increasing gate voltages which contradicts the respective gate dependence of the simulation (Figure~\ref{fig:Sim_LongLifetimes}f). Moreover, the gate dependent spin diffusion length in Figure~\ref{fig:Sim_LongLifetimes}f shows a clear underestimation at larger gate voltages which is caused by an underestimation of $D_\text{s}$ (see difference between solid line and data points in Figure~\ref{fig:Sim_LongLifetimes}f).
Thus, it would be expected that $D_\text{s}$ deviates from $D_\text{c}$ in the experiment which is not the case (see Ref.~\cite{doi:10.1021/acs.nanolett.6b00497}). 
Therefore, we exclude that the finite flake size is responsible for the measured long spin lifetimes.

\section{Spatially-varying spin transport parameter}

\begin{figure*}[tbh]
	\centering
		\includegraphics{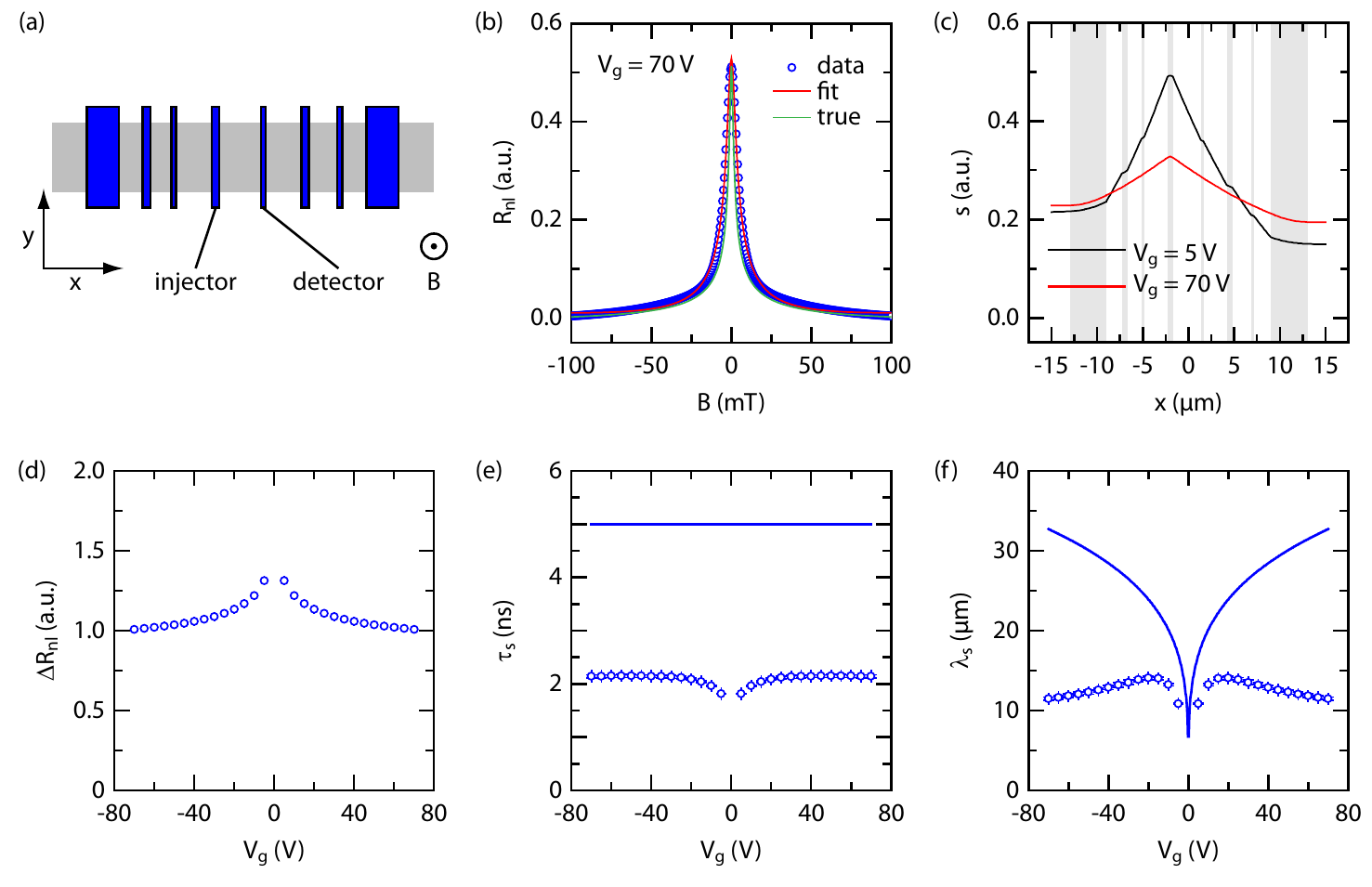}
	\caption{Simulation in the case of contact-induced dephasing. (a) Geometry of the spin transport device with assigned spin injection and spin detection electrodes. The blue areas depict the contact-covered areas where the spin lifetime of graphene is reduced to $\tau_\text{s}=\unit[0.5]{ns}$ and the diffusion coefficient is assumed to be gate independent and is fixed at $D_\text{s}=\unit[0.2]{m^2/s}$. For the graphene in between the electrodes we again set the spin lifetime to $\tau_\text{s}=\unit[5]{ns}$ and the electron mobility to $\mu=\unit[20,000]{cm^2/(Vs)}$ which means that $D_\text{s}$ changes which the applied gate voltage, i.e. induced charge carrier density. (b) Simulated Hanle curve for $V_\text{g}=\unit[70]{V}$. Next to the simulated data points also the fit (red curve) and the expected Hanle curve (green curve) for the genuine spin transport parameters are included. (c) Spatial variation of the spin accumulation for $V_\text{g} = \unit[5]{V}$ and $V_\text{g} = \unit[70]{V}$. The positions of the contacts are indicated by the shaded areas. (d-f) Fitted spin transport parameters as a function of gate voltage: (d) The magnitude of the spin signal $\Delta R_\text{nl}$, (e) the spin lifetime $\tau_\text{s}$, and (f) the spin diffusion length $\lambda_\text{s}$. All data were extracted from the simulated Hanle curves. The solid lines in (e) and (f) show the respective values which were used as parameter for the simulation.}
	\label{fig:Sim_ContactInducedDephasing}
\end{figure*}

Contact-induced spin dephasing has been discussed to be a dominant dephasing mechanism in graphene \cite{PhysRevB.88.161405,PhysRevB.90.165403,2053-1583-2-2-024001,Amamou2016,PhysRevApplied.6.054015}. Therefore, it is interesting to evaluate the gate dependent spin transport parameters in such a non-uniform system where we spatially vary the spin transport properties. For the contact-covered areas (blue areas in Figure~\ref{fig:Sim_ContactInducedDephasing}a) we assume a spin lifetime of $\tau_\text{s}=\unit[0.5]{ns}$ and a spin diffusion coefficient of $D_\text{s}=\unit[0.2]{m^2/s}$. The latter value equals the diffusion coefficient at $V_\text{g} = \unit[70]{V}$ for a device mobility of $\mu=\unit[20,000]{cm^2/(Vs)}$. In this case the resulting charge carrier density matches the one calculated from the contact-induced doping density which was extracted from confocal Raman spectroscopy \cite{Droegeler2014}. Because of shielding of the gate electric field by the electrodes and/or Fermi level pinning \cite{PhysRevLett.101.026803,doi:10.1021/nl300266p}, both the spin lifetime and the spin diffusion coefficient are assumed to be independent on the gate voltage in all contact areas in Figure~\ref{fig:Sim_ContactInducedDephasing}a (blue areas). For the bare parts of the graphene flake we use the same spin lifetime of $\tau_\text{s}=\unit[5]{ns}$ and the same mobility of $\mu=\unit[20,000]{cm^2/(Vs)}$ as above. In this case the resulting diffusion coefficient is gate dependent.

\begin{figure*}[tbh]
	\centering
		\includegraphics{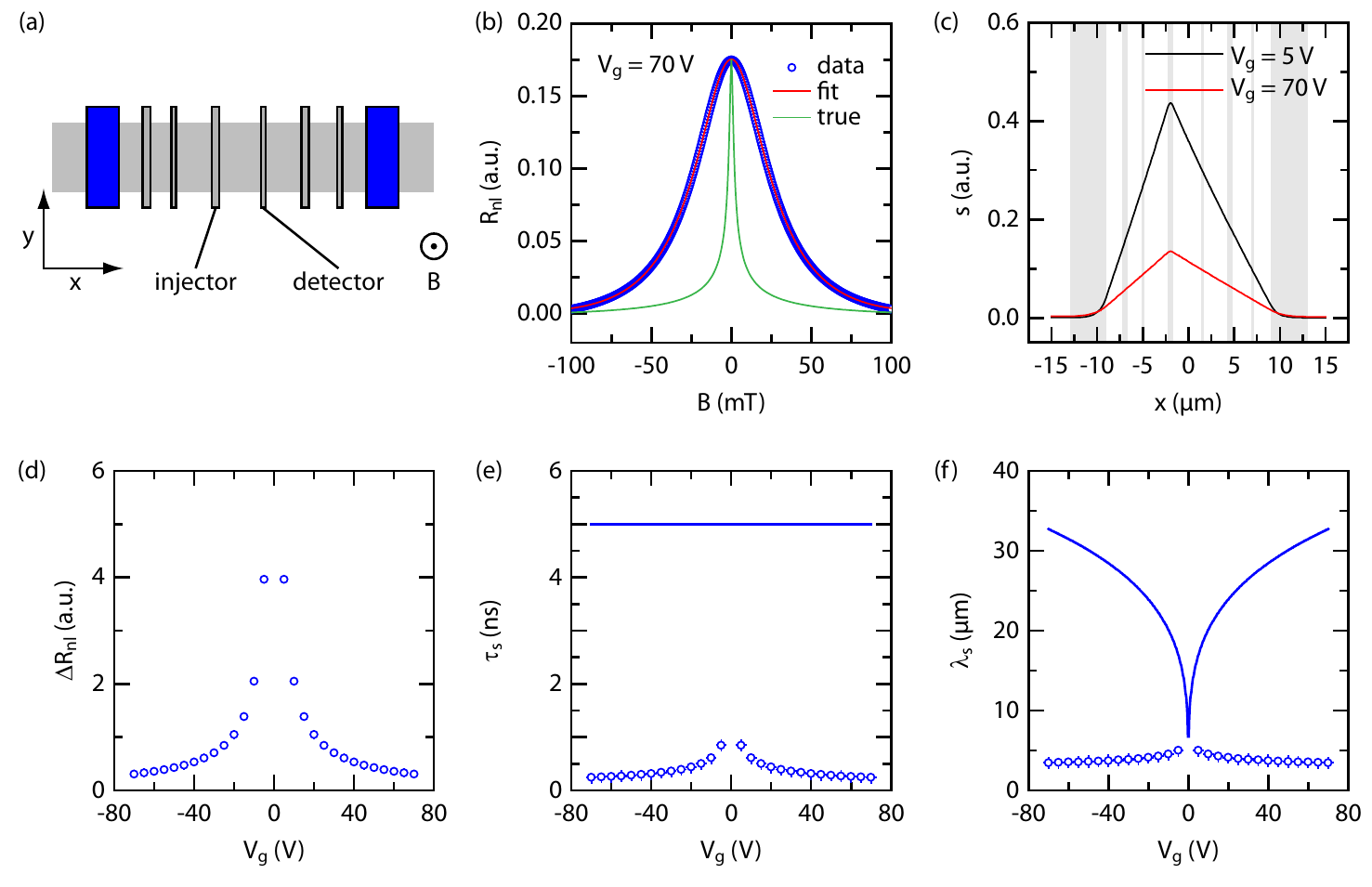}
	\caption{Simulations in case of gold reference electrodes which can act as a spin sink. (a) Geometry of the spin transport device with assigned spin injection and spin detection electrodes. The blue areas depict the position of the reference electrodes where the spin lifetime is reduced to $\tau_\text{s}=\unit[10]{ps}$. For the remaining graphene flake we set a spin lifetime of $\tau_\text{s}=\unit[5]{ns}$. The diffusion coefficient is assumed to be gate tunable and is calculated from the induced charge carrier density and an electron mobility of $\mu=\unit[20,000]{cm^2/(Vs)}$ for the entire flake as before. (b) Simulated Hanle curve for $V_\text{g}=\unit[70]{V}$. Next to the simulated data points also the fit (red curve) and the expected Hanle curve (green curve) for the genuine spin transport parameters are included. (c) Spatial variation of the spin accumulation for $V_\text{g} = \unit[5]{V}$ and $V_\text{g} = \unit[70]{V}$. The positions of the contacts are indicated by the shaded areas. (d-f) Spin transport parameters as a function of gate voltage: (d) The magnitude of the spin signal $\Delta R_\text{nl}$, (e) the spin lifetime $\tau_\text{s}$, and (f) the spin diffusion length $\lambda_\text{s}$. All data were extracted from the simulated Hanle curves. The solid lines in (e) and (f) show the respective values which were used as parameter for the simulations.}
 	\label{fig:Sim_GoldContacts}
\end{figure*}

We next discuss the the gate dependent spin transport properties (shown in Figures~\ref{fig:Sim_ContactInducedDephasing}d-f) which we extract from fitting the simulated Hanle curves to the simplified equation which assumes uniform spin lifetimes and spin diffusion coefficients. An example of a simulated Hanle curve is shown in Figure~\ref{fig:Sim_ContactInducedDephasing}b for $V_\text{g}=\unit[70]{V}$. Additionally, the respective fit (red curve) and the expected Hanle curve (green curve) are shown. For the amplitude of the spin signal $\Delta R_\text{nl}$ there is a decrease for increasing gate voltages as it is expected for higher charge carrier mobilities consistent with the results in the previous section. For the spin lifetime a small increase for low gate voltages can be observed before it approaches a value of around $\tau_\text{s}=\unit[2.2]{ns}$. This value roughly corresponds to the average of the spin dephasing rates ($1/\tau_\text{s}$) which is weighted by their contribution to the transport length $L_\text{cc}$ given by the center-to-center distance between the injector and detector contacts. In this case, the actual spin lifetime is underestimated by more than a factor of two. However, the increase in spin lifetime for low gate voltages is similar to what is observed in experiments \cite{Droegeler2014,doi:10.1021/acs.nanolett.6b00497,PhysRevLett.113.086602,PhysRevB.88.161405,PhysRevB.90.165403,PhysRevB.80.241403,PhysRevLett.107.047207,doi:10.1021/nl301567n,PhysRevLett.107.047206,Avsar2011,PSSB201552418}. But the relative change in amplitude in the simulation is significantly smaller. For the spin diffusion length it can be seen that the values first increase for low gate voltages and thereafter decrease for larger gate voltages. This behaviour can be understood when considering the spatial distribution of the respective spin accumulations (Figure~\ref{fig:Sim_ContactInducedDephasing}c). For $V_\text{g} = \unit[5]{V}$ distinct kinks can be observed in the regions of the contacts which are caused by the larger spin diffusion coefficient. Nevertheless, the overall measured diffusion coefficient is mainly determined by the contact-free parts of the graphene which lead to an increase of the diffusion length. For $V_\text{g}=\unit[70]{V}$ the kinks in the spin accumulation are not visible anymore. Here, the diffusion coefficients of both regions (contact-covered and bare graphene) become similar. However, the extracted average diffusion is suppressed due to the finite flake size as seen before.

\section{Metallic reference electrodes}

In the previous section it was shown that an enhanced spin dephasing underneath the contacts leads to an overall reduced spin lifetime. The next simulation investigates the role of outer metallic reference electrodes such as gold which has been used in some experiments \cite{PhysRevLett.107.047207,doi:10.1021/nl301567n,PhysRevLett.109.186604}. Due to the large spin orbit interaction in gold and the absence of an injection and detection barrier between the gold and graphene, these electrodes are expected to act as a spin sink. As a result, the spin lifetimes in gold-covered graphene regions are expected to be very short. Therefore, the spins most likely will not reach the end of the graphene flake and the spin lifetime will not be overestimated. However, it is not clear how the actual measurements will be influenced. For the simulation we assume a spin lifetime of $\tau_\text{s}=\unit[10]{ps}$ for the gold contacts while for the bare graphene again $\tau_\text{s}=\unit[5]{ns}$ is used. Additionally, a gate dependent diffusion coefficient is used for the entire flake and a mobility of $\mu=\unit[20,000]{cm^2/(Vs)}$ is assumed.

The respective fit results are plotted in Figures~\ref{fig:Sim_GoldContacts}d-f. An example of a simulated Hanle curve for $V_\text{g}=\unit[70]{V}$, the respective fit (red curve) and the expected Hanle curve (green curve) are shown in Figure~\ref{fig:Sim_GoldContacts}b. Similar to the previous cases, the amplitude of the spin signal decreases towards higher gate voltages but this time the effect is more pronounced. This can be explained by the fact that for larger diffusion coefficients (higher gate voltages) more spins reach the gold electrodes and dephase. Hence, the spin sink effect is more effective at large gate voltages. A similar dependence can also be seen in the gate dependent spin lifetime. Here, the lifetime is only on the order of $\tau_\text{s}=\unit[250]{ps}$ at large gate voltages which is much smaller than the assumed value of $\tau_\text{s}=\unit[5]{ns}$. This means that also non-magnetic contact regions which are not part of the studied spin transport channel and which additionally exhibit short spin lifetimes can lead to tremendous underestimation of the real spin lifetime. Additionally, also the spin diffusion lengths exhibit the largest values at low gate voltages and a drop towards higher gate voltages to an almost constant value (Figure~\ref{fig:Sim_GoldContacts}f). When investigating the spatial distribution of the spin accumulation in Figure~\ref{fig:Sim_GoldContacts}c, it becomes obvious that, due to the short spin lifetime in the graphene areas of the gold contacts, the spin accumulation is basically forced to be zero, i.e. the gold contacts act as a spin sink. This of course changes the spatial gradient of the spin accumulation and, therefore, the diffusion process. To enhance the spin transport properties in future devices, our results suggest using only decoupled non-magnetic reference electrodes by, for example, adding additional injection and detection barriers between the non-magnetic metal and graphene which is expected to maintain the spin polarization.

\section{Conclusion}

In summary we have demonstrated that only in case of high quality graphene with uniform spin transport parameters it is possible to overestimate the experimentally extracted spin lifetimes from Hanle spin precession curves. In contrast, if one graphene region in a non-local spin valve device exhibits a short spin lifetime by, for example, contact-induced spin dephasing, this will also effect the spin lifetime extracted in any other region of the device which become underestimated. This impact is almost independent of the particular location of the non-uniformity as long as it is positioned within the spin diffusion length relative to the spin injection electrode. However, none of the simulations showed a significant increase of the spin lifetime with increasing gate voltage indicating that this dependence is not directly linked to the non-uniformities but rather attributed to a spin dephasing process in contact-free graphene parts of the device. This does not exclude the possibility that the overall spin lifetime is still influenced by contact-induced spin dephasing. We conclude that an ideal graphene-based spin transport should have a long graphene flake, i.e. exceeding $\unit[100]{\mu m}$, to avoid back reflection of the spins. The non-magnetic reference electrodes should be properly decoupled from the graphene by using insulating barriers. These barriers, which are also used for the spin injection and spin detection, need to be homogeneous and thick enough in order to not influence the transport properties of the graphene flake. Additionally, the length of the transport region can be enhanced in order to minimize potential spin scattering from the contacts.

%
%
%
\section{acknowledgement}
We acknowledge funding from the European Union Seventh Framework Programme under Grant Agreement No. 696656 Graphene Flagship and the Deutsche Forschungsgemeinschaft (BE 2441/9-1).

\end{document}